\documentclass[11pt]{article}
\usepackage{acl2015}
\usepackage{times}
\usepackage{url}
\usepackage{latexsym}
\usepackage{graphics} 
\usepackage{epsfig} 
\usepackage{mathptmx} 
\usepackage{times} 
\usepackage{amsmath} 
\usepackage{amssymb}  
\usepackage{caption}
\usepackage{subcaption}
\usepackage{float}
\usepackage{multirow}
\usepackage{pbox}
\usepackage{breqn}
\usepackage{subcaption}
\usepackage{graphicx}

\title{Autoencoder Based Architecture for Fast \& Real Time Audio Style Transfer}

\author{Dhruv Ramani, Samarjit Karmakar, Anirban Panda, Asad Ahmed, Pratham Tangri
\\
Nevronas
\\
National Institute of Technology, Warangal, India\\
{\tt \{dhruvramani98, karmakar.samarjit, ahmed.asad19, prathamtangri2015\}@gmail.com,} \\
{\tt anirban.panda@ieee.org}
}

\date{}

\begin{document}
\maketitle
\begin{abstract}
  Recently, there has been great interest in the field of audio style transfer, where a stylized audio is generated by imposing the style of a reference audio on the content of a target audio. We improve on the current approaches which use neural networks to extract the content and the style of the audio signal and propose a new autoencoder based architecture for the task. This network generates a stylized audio for a content audio in a single forward pass. The proposed network architecture proves to be advantageous over the quality of audio produced and the time taken to train the network. The network is experimented on speech signals to confirm the validity of our proposal.  
\end{abstract}

\section{Introduction}

The task of artistic style transfer has been far studied and implemented for generating stylized images. It provides an insight that the content and style representation of visual imagery are separable. Style transfer in images can be explained as imposing the style extracted from a reference image onto the content of a target image. The seminal works of Gatys et al~\shortcite{Gatys:16} and Johnson et al~\shortcite{Johnson:16}, shows the usage of convolutional neural networks (CNNs) for the task.
CNNs prove to be advantageous for the task because of the representations learned by it in the deeper layers. These deep features can be used to represent the content and the style of an image, separately. This has led to a great increase of research in the area of style transfer which incorporates neural networks to transfer the "style" of an image (eg. a painting) to another (eg. a photograph). 

The task of audio style transfer is recently gaining popularity as an area of research because of its wide applications in audio editing and sound generation. The meaning of style and content for an audio signal is different as compared to an image.  The current consensus implies that style refers to the speaker's identity, accent, intonation, and the content refers to the linguistic information enccoded in it like phonemes and word. Over time, various methods have been proposed which involve usage of models used for image style transfer on audio. This involves converting the raw audio into a spectrogram and using neural networks to extract the required features. Further, we generate waveforms which match the high level network activations from a content signal while simultaneously matching low level statistics computed from lower level activations from a style signal. In this paper, we propose a new and a novel architecture which uses similar approaches to stylize an audio. But unlike previous proposed methods, our proposed architecture stylizes an audio in a single network pass and is thus extremely useful for real time audio style transfer. The network architecture is carefully crafted to ensure faster training time and low computational usage. 
In this paper, we explore previous methods which have been proposed for artistic style transfer in images and audio, propose a new architecture and analyze it's performance.

\begin{figure*}
\centering
\includegraphics[height=0.14\textheight,width=1.00\textwidth]{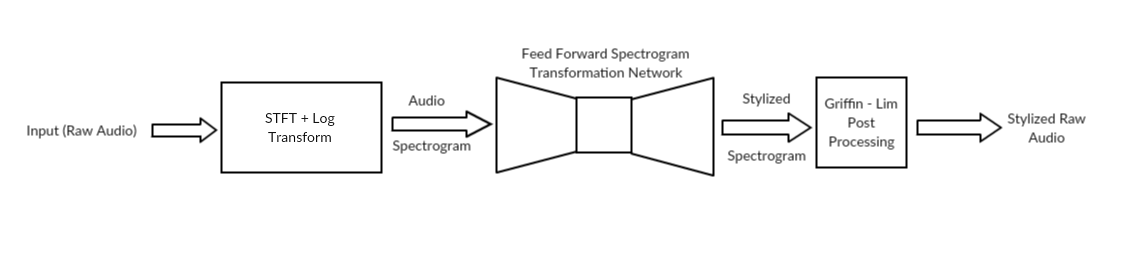}
\caption{A framework for audio style transfer using a single convolutional autoencoder, trained on spectrograms of speech signals and a single style signal is used to generate stylized audio. The signal is pre-processed by applying Short Time Fourier Transform (STFT) on the raw input audio to generate an audio spectrogram. This spectrogram is passed through the transformation network to generate the stylized spectrogram. For retrieveing the audio back from the spectrogram, Griffin-Lim algorithm is used to convert the stylized spectrogram to the required stylized audio.}
\label{figure1}
\end{figure*}

\section{Related Work}

The work of Gatys et al~\shortcite{Gatys:16} shows the advantageous use of convolutional neural networks (CNNs) for stylizing the target image. In it, the content of an image is conceptualized as high level representation gained from the deeper layers of a CNN trained for image classification. The style representation of an image is taken as a linear combination of the gram matrix of the feature maps of different layers of the same network.

Let the filter response tensor of the $l^{th}$ layer of the network be $F_l(S)$, where $S$ is the style image, then the gram matrix representation of this layer is given as:
\begin{equation}
    Gram(F_l(S)) = F_l(S) \cdot F_l(S)^T
\end{equation}

The above method was slow and had to be iteratively optimized for each content-style pair to obtain a single stylized output. Johnson et al~\shortcite{Johnson:16} proposed a transformation network based architecture. In this, they train a transformation network on content images, imposing the style of a single style image to generate the stylized output. The network learns a mapping from the content images to stylized images, which are biased towards a single style image. The stylized outputs are generated in a single forward pass of the network, hence, this method has aptly been named fast neural style transfer and is extremely useful for real time style transfer applications. 

The loss for the network is given by taking into account a measure of content and style similar to what was defined by Gatys et al~\shortcite{Gatys:16}. A VGG network which had been pre-trained for image classification is used for extracting content and style from the respective images. The content and style representations are obtained in the same way as mentioned before. The loss function captures both high and low level information of the image. The method also accounts for total variation loss which improves spatial smoothness in the output image. The total loss given by,
\begin{equation}
    l_{total} = \alpha l_{content}(Y, C) + \beta l_{style}(Y, S) + \gamma l_{tv}
\end{equation}
where $Y$ is the output of the transformation network, $C$ is the content image and $S$ is the style image. This loss is minimized by backpropagation using Stochastic Gradient Descent (SGD) as an optimizer.

The work of Ulyanov~\shortcite{Ulyanov:16} on audio
style transfer, used a similar optimization framework as Gatys et al~\shortcite{Gatys:16}, but unlike Gatys et al, they did not use a deep pre-trained neural network, instead opting for a shallow network (A single layer with 4096 random filter). In this, the output spectrogram is initialized as a random noise. This is then optimized using the model till the loss between features taken from content and style audio, and the output audio, is minimized. However the results of this were limited and preservation of content and style was not to a high degree. The work by Grinstein et al~\shortcite{Grinstein:18} utilized this meaning, however, the work only considered style loss since the original audio itself was being modified, the content loss wasn't considered.

Audio style transfer has been experimented on iterative optimization based approaches using neural networks, opening up the scope for research in real time approaches which can generate the stylized audio in a single forward pass of a feed-forward neural network.

\begin{figure*}
\centering
\includegraphics[height=0.24\textheight,width=0.70\textwidth]{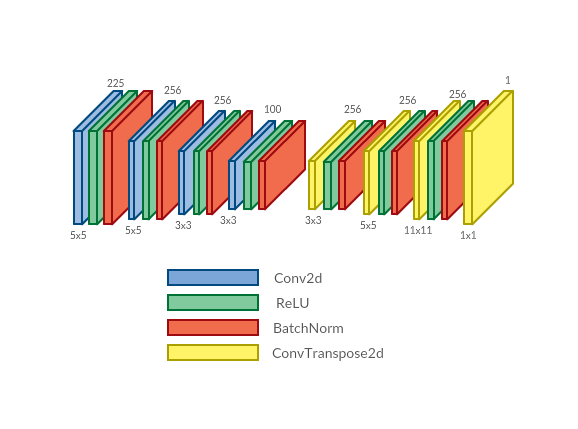}
\caption{An autoencoder based architecture for the transformation network and the loss network. The number of filters and the kernel size for each layer appear above and below the layer blocks respectively in the diagram.}
\label{figure2}
\end{figure*}

\section{Problem Definition and Formulation}

We aim to solve the problem of neural audio style transfer using a transformation network and a loss network.

Given a content audio $A$ and a style audio $B$, our task is to find the audio $\hat{Z}$ which satisfies the equation,
\begin{equation}
    \hat{Z} = \underset{Z}{\mathrm{argmin}}(\alpha ||C(Z) - C(A)||^2 + \beta ||S(Z) - S(B)||^2)
\end{equation}
Here, $C$ represents content of an audio and $S$ represents the style of an audio. $\alpha$ and $\beta$ are parameters which signify the amount of content or style we require in our output audio $\hat{Z}$. A higher value of $\alpha$ over $\beta$ would result in a predominance of content in the audio $\hat{Z}$.

\section{Our Proposed Architecture}
The general framework we propose to adopt for the purpose of real time audio style transfer is illustrated in Figure~\ref{figure1}.

The raw speech signal contains all the information in the temporal domain. The signal is pre-processed so that it can be reconstructed using Griffin-Lim algorithm~\cite{griffin:84} later. Initially, Short Term Fourier Transform (STFT) is applied to bring the raw audio from time domain to frequency domain. This helps us to understand the frequency range at which the signal emphasizes. This relative emphasis helps in shaping the high level features like phoneme or emotion. The frequency domain signal is then converted into magnitude-spectral domain by taking the magnitude of the result of STFT. The magnitude is chosen over the phase as it provides higher and richer information about the high level features and helps in easier reconstruction of the signal. The spectrum of this signal is obtained by taking log of magnitude with time as the horizontal axis and frequency as the vertical axis. The frequency is transformed to the log scale to visualize the features related to human perception of natural sound. The obtained spectrum of the speech signal is known as an audio spectrogram. This can be thought of an image representation of the audio signal, except the fact that translation along frequency axis can change high-level features like emotion and doesn’t change features like words spoken. 
Mathematically, let $x(t)$ be the input raw audio signal. The spectrogram for the signal $x(t)$ for a window function $w$ is given by,
\begin{equation}
    Spec\{x(t), w\}=log_{e}(|STFT\{x(t), w\}|)
\end{equation}
where, the $STFT$ function is given by,
\begin{equation}
    STFT\{x(t), w\}= \int_{-\infty}^{+\infty}x(t)w(t-\tau)e^{-j\omega t}dt
\end{equation}

A network architecture similar to that proposed by Johnson et al~\shortcite{Johnson:16} is adopted. A transformation network parameterized by $\lambda_{trans}$ is utilized to find a mapping from an input space of content audio spectrogram to an output space of stylized audio spectrogram. To calculate the loss, a loss network parameterized by $\lambda_{loss}$ is used to extract the content from the respective spectograms. Subsequently, the same loss network is used to extract the style from the spectograms. The loss network is pretrained to extract a hierarchy of representation from the audio spectrogram, incorporating both low level features and high level features.

\begin{figure*}
\centering
\includegraphics[height=0.25\textheight,width=1.00\textwidth]{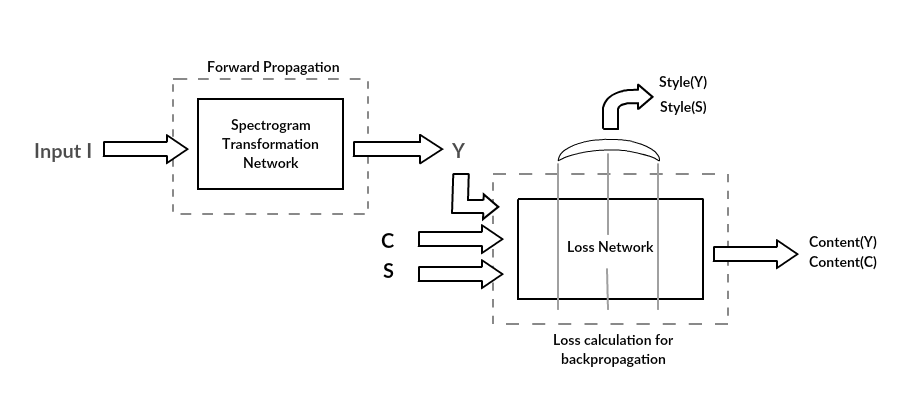}
\caption{A framework for training the spectrogram transformation network ($STN$) and loss calculation for the purpose of backpropagation of $STN$.}
\label{figure3}
\end{figure*}

\subsection{Loss Network}
We adopt an encoder-decoder architecture~\cite{auto:17}, illustrated in Figure~\ref{figure2}, for the loss network. It consists of 4 convolutional layers and 4 transposed  convolutional layers. ReLU non-linearity followed by Batch Normalization is applied to all the layers except the last layer. The network is treated as an autoencoder. The network compresses the input spectrogram to lower dimensional latent space and further tries to reconstruct the same input. As a consequence of this process, the encoder part of the autoencoder learns to capture high level features of the input spectrogram and is able to represent the same in lower dimensions, called as latent embedding.. The decoder part is used to reconstruct the spectrogram from the latent embedding. This is then optimized with backpropagation to ensure that the reconstructed spectrogram is a similar to the input. 

The latent embedding feature activation map is used to model content of a spectrogram and the linear combination gram matrix of feature activation maps of first, second and third convolutional layers is used to model style of a spectrogram.

\subsection{Transformation Network}
We use the same encoder-decoder architecture, illustrated in Figure~\ref{figure2}, for the tranformation network. Instead of training the transformation network from scratch, pretrained weights from the loss network are used. This approach is impressive as it utilizes the weights of a pretrained neural network which has learned the distribution of audio spectrograms, and therefore does not require re-learning of the content representation. The network is only optimized to accommodate the low level features of a single spectrogram of a given style which need not be related to the samples which had been used to train the network. This makes the training process faster comparatively while ensuring proper results in under one epoch of training. Using the same architecture also ensures homogeneity within the dimensions of the spectrogram.  A framework for training this network is illustrated in Figure\ref{figure3}.

An input spectrogram ($I$) is passed through spectrogram transformation network ($STN$) to generate an output spectrogram ($Y$). The weights and biases of the loss network ($LN$) are frozen. $LN$ is used for calculation of content of $Y$ ($Con(Y)$), style of $Y$ ($Sty(Y)$), content of $C$ ($Con(C)$) and style of $S$ ($Sty(S)$). Since we want to preserve the content of the input spectrogram, here $I$ and $C$ will be the same spectrogram. The loss is calculated as:

\begin{dmath}
loss = \alpha \times MSE(Con(Y), Con(C)) + \beta \times MSE(Sty(Y), Sty(S))
\label{eq1}
\end{dmath}
This loss is minimized by optimizing the weights and biases of $STN$ using backpropagation.

\section{Experiments}
To train the proposed architecture on speech signals, we use the publicly available CSTR VCTK Corpus~\cite{vctk:12}. The VCTK corpus provides text labels for the speech and is widely used in speech to text synthesis. However, since we employ an autoencoder based architecture, the text labels aren't used. The corpus contains clean speech from 109 speakers which read out 400 sentences, the majority of which have British accents. We down sampled the audio to 16 kHz for our convenience.

To convert the raw audio signal to a spectrogram, we apply Short Term Fourier Transform (STFT) to bring the speech utterance to a frequency domain. We then apply $log$ to the magnitude of the signal to convert it into a magnitude-spectral domain. The signal is now in the form of an audio spectrogram. The loss network is trained on the spectrograms of audios from the VCTK corpus. We optimize the weights and biases using backpropgation to minimize the mean squared error between the reconstructed spectrogram and the input provided. The loss is minimized using Adam~\cite{adam:15} optimizer with a learning rate of $10^{-3}$, weight decay of $10^{-4}$ and batch size as $24$. The activations of this network are used to represent the content and style of a signal, separately. 

While training the transformation network, the weights and biases of the loss network are kept frozen. We use pre-trained weights of loss network for training the transformation network for faster learning. The pre-processed signals from the corpus are used as content samples and a single style sample is used for training the transformation network. As a result, the network gets biased towards generating stylized output spectrograms pertaining to a single style for any content spectrogram. The loss in Equation~\ref{eq1} is minimized using Adam~\cite{adam:15} optimizer, with learning rate of $10^{-3}$, $\beta_{1}$ as 0.999 and $\beta_{2}$ as 0.99. The value of $\alpha$ in the loss function is taken as 100 and $\beta$ is taken as $10^4$. These values had been chosen after excessive experimentation.

The models have been implemented using the PyTorch deep learning framework and trained on a single Nvidia GTX 1070 Ti GPU. 
\begin{figure}

\begin{subfigure}{\linewidth}
\includegraphics[trim={15cm 0 4cm 0},clip, height=0.05\textheight,width=1.00\textwidth]{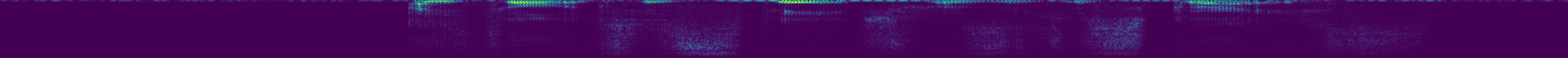}
\caption{The original content audio spectrogram.}
\label{figure4}
\end{subfigure}

\begin{subfigure}{\linewidth}

\includegraphics[trim={68cm 0 70cm 0},clip,height=0.05\textheight,width=1.00\textwidth]{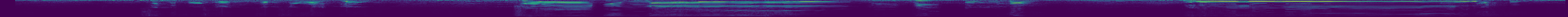}
\caption{The style audio spectrogram.}
\label{figure5}
\end{subfigure}

\begin{subfigure}{\linewidth}
\includegraphics[trim={15cm 0 4cm 0},clip, height=0.05\textheight,width=1.00\textwidth]{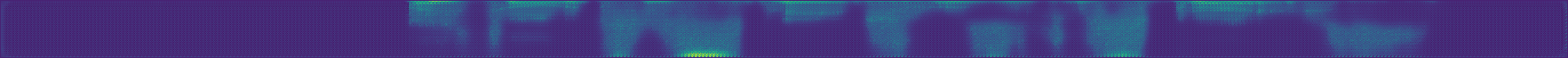}
\caption{The stylized output audio spectrogram.}
\label{figure6}
\end{subfigure}
\label{figure7}
\caption{Audio spectrograms in $log$ magnitude-spectral domain}
\end{figure}

\section{Results}
The key finding is that the low level statistical information from a style audio spectrogram, kept constant while the spectrogram transformation network is trained, can be transferred to a target spectrogram, while preserving the content of it, in a single forward pass of the network, while testing. The qualitative results in the form of spectrograms of the content, style utterance and stylized output generated using this architecture are shown in Figure~\ref{figure4}, ~\ref{figure5} and ~\ref{figure6}. From the spectrograms, it can be observed that the content is retained, however the output contains very different properties such as pitch, accent, etc. The texture of the style audio spectrogram is present in the output audio spectrogram, whereas the content, defined by the lightly shaded regions within the dark background in the original content audio spectrogram, is retained. The stylized output audio spectrogram may be converted into a raw audio signal by post-processing using Griffin-Lim algorithm for further auditory analysis to support this claim.

\section{Conclusion}
In this work, we propose a new architecture for real time audio style transfer. We experimented and evaluated the proposed model on several speech utterances and the model is found to show promising results. Since a single transformation network is trained for stylizing the content audio on a specific style audio, the style audio isn't needed during testing. In future work, more conducive research and experimentation on the meaning and representation of "style" in audio will result in the discovery of features which will help generate various forms of the stylized audio. Moreover, as the loss network is separated from the transformation network, we may incorporate several other features for calculation of loss, such as accent or music, to have only specific features transferred to the generated audio and preserving several other features.

\section*{Acknowledgments}
We would like to thank Innovation Garage, NIT Warangal for their invaluable help in providing us with necessary computing capabilities which made our research possible.


\begin{thebibliography}{}
\bibitem[\protect\citename{Gatys et al.}2016]{Gatys:16}
Leon A. Gatys, AlexandS. Ecker, Matthias Bethge.
\newblock 2016.
\newblock {\em Image Style Transfer Using Convolutional Neural Networks}.
\newblock Proceedings of IEEE Conference on Computer Vision and Pattern Recognition (CVPR) 2016.

\bibitem[\protect\citename{Johnson et al.}2016]{Johnson:16}
Justin Johnson, Alexandre Alahi, Li Fei-Fei.
\newblock 2016.
\newblock {\em 
Perceptual Losses for Real-Time Style Transfer and Super-Resolution
}.
\newblock Proceedings of European Conference on Computer Vision (ECCV) 2016.

\bibitem[\protect\citename{Grinstein et al.}2018]{Grinstein:18}
Eric Grinstein, Ngoc Duong, Alexey Ozerov, Patrick Pérez.
\newblock 2018.
\newblock {\em Audio style transfer}.
\newblock Proceedings of IEEE International Conference on Acoustics, Speech and Signal Processing (ICASSP) 2018.

\bibitem[\protect\citename{Ulyanov et al.}2016]{Ulyanov:16}
Dmitry Ulyanov.
\newblock 2016.
\newblock {\em Audio Texture Synthesis and Style Transfer}.
\newblock https://dmitryulyanov.github.io/audio-texture-synthesis-and-style-transfer/.

\bibitem[\protect\citename{Simonyan and Zisserman}2015]{vgg:15}
Karen Simonyan, Andrew Zisserman.
\newblock 2015.
\newblock {\em Very Deep Convolutional Networks for Large-Scale Image Recognition}.
\newblock Proceedings of International Conference on Learning Representations (ICLR) 2015.

\bibitem[\protect\citename{Deng et al.}2009]{imagenet:09}
Jia Deng, Wei Dong, Richard Socher, Li-Jia Li, Kai Li, Li Fei-Fei.
\newblock 2009.
\newblock {\em ImageNet: A Large-Scale Hierarchical Image Database}.
\newblock Proceedings of IEEE Conference on Computer Vision and Pattern Recognition (CVPR) 2009.

\bibitem[\protect\citename{Kingma and Ba}2015]{adam:15}
Diederik P. Kingma, Jimmy Ba.
\newblock 2015.
\newblock {\em Adam: A Method for Stochastic Optimization}.
\newblock Proceedings of International Conference for Learning Representations (ICLR) 2015.

\bibitem[\protect\citename{Yamagishi and Junichi}2012]{vctk:12}
Yamagishi, Junichi.
\newblock 2012.
\newblock{\em English multi-speaker corpus for CSTR voice cloning toolkit}.
\newblock http://homepages.inf.ed.ac.uk/jyamagis/page3/
page58/page58.html .

\bibitem[\protect\citename{Perez et al.}2017]{auto:17}
Anthony Perez, Chris Proctor, Archa Jain.
\newblock 2017.
\newblock{\em Style Transfer for Prosodic Speech}.
\newblock http://web.stanford.edu/class/cs224s/reports/  \newline Anthony\_Perez.pdf .

\bibitem[\protect\citename{Griffin and Lim}1984]{griffin:84}
D. Griffin, Jae Lim.
\newblock 1984.
\newblock{\em Signal estimation from modified short-time Fourier transform}.
\newblock  IEEE Transactions on Acoustics, Speech, and Signal Processing 1984.

\end{thebibliography}
\end{document}